\documentclass[prd,aps,preprint,nofootinbib,amssymb,eqsecnum,showkeys]{revtex4}
\usepackage{verbatim,graphics,graphicx,color,slashed}
\usepackage{ulem} 
\usepackage{hyperref}
\usepackage{changepage}

\def\lsim{\mathrel {\vcenter {\baselineskip 0pt \kern 0pt
    \hbox{$<$} \kern 0pt \hbox{$\sim$} }}}
\def\gsim{\mathrel {\vcenter {\baselineskip 0pt \kern 0pt
    \hbox{$>$} \kern 0pt \hbox{$\sim$} }}}
\def\slashchar#1{\setbox0=\hbox{$#1$}           
 \dimen0=\wd0                                 
  \setbox1=\hbox{/} \dimen1=\wd1               
\ifdim\dimen0>\dimen1                        
  \rlap{\hbox to \dimen0{\hfil/\hfil}}      
  #1                                        
  \else                                        
 \rlap{\hbox to \dimen1{\hfil$#1$\hfil}}   
   /                                         
  \fi}                                         %
\def\cpto{\mathrel {\vcenter {\baselineskip 0pt \kern 0pt
    \hbox{$CP$} \kern 0pt \hbox{$\longrightarrow$} }}}
\def\cptof{\mathrel {\vcenter {\baselineskip 0pt \kern 0pt
    \hbox{$~CP$} \kern 0pt \hbox{$\longleftrightarrow$} }}}

\begin{document}

\baselineskip=15pt

\preprint{}

\title{Large $SU(3)$ breaking effects and  CP violation in $B^+ $ decays\\ into three charged octet pseudoscalar mesons}

\author{Dong Xu${}^{1}$\footnote{xudong1104@gmail.com}}
\author{Guan-Nan Li$^{1}$\footnote{lgn198741@126.com}}
\author{Xiao-Gang He$^{1,2,3}$\footnote{hexg@phys.ntu.edu.tw}}
\affiliation{${}^{1}$INPAC, SKLPPC and Department of Physics,
Shanghai Jiao Tong University, Shanghai, China}
\affiliation{${}^{2}$Physics Division, National Center for
Theoretical Sciences, Department of Physics, National Tsing Hua
University, Hsinchu, Taiwan} \affiliation{${}^{3}$CTS, CASTS and
Department of Physics, National Taiwan University, Taipei, Taiwan}

\date{\today $\vphantom{\bigg|_{\bigg|}^|}$}

\date{\today}

\vskip 1cm
\begin{abstract}
The LHCb collaboration has recently reported evidence for non-zero CP asymmetries in $B^+$ decays into  $\pi^+ K^+ K^-,\; \pi^+\pi^+\pi^-,\; K^+ K^+ K^- $ and $K^+\pi^+\pi^-$. The branching ratios for these decays have also been measured with different values ranging from $5\times 10^{-6}$ to $51\times 10^{-6}$. If flavor $SU(3)$ symmetry is a good symmetry for $B$ decays,  in the case that the dominant amplitude is momentum independent  it is expected that branching ratios $Br$ and CP violating rate differences $\Delta_{CP} = \Gamma - \overline{\Gamma}$ satisfy, $Br(\pi^+\pi^+\pi^-) = 2Br(\pi^+ K^+ K^-)$,
$Br(K^+K^+K^-) = 2 Br(K^+\pi^+\pi^-)$, and $\Delta_{CP}(\pi^+\pi^+\pi^-) = 2\Delta_{CP}(\pi^+ K^+K^-) = -
\Delta_{CP}(K^+K^+K^-) = -2\Delta_{CP}(K^+\pi^+\pi^-)$. The experimental data do not exhibit the expected pattern for the branching ratios. The rate differences for $B^+\to \pi^+\pi^+\pi^-$ and $B^+\to K^+ K^+ K^-$ satisfy the relation between $\Delta S =0$ and $\Delta S=1$ well, but the other two do not, with the CP asymmetries having different signs than expected. In this work we study how to including momentum dependent  and also $SU(3)$ breaking effects on these decays to explain experimental data. We find that only including lowest order derivative terms, in the $SU(3)$ limit, the decay patterns cannot be explained. Large $SU(3)$ breaking effects are needed to explain the data.
\end{abstract}

\pacs{PACS numbers: }

\maketitle

\section{Introduction}

Decays involving a heavy b-quark have been a subject of very active research in the past decade and continue to be so at present. Data on $B$ decays from SLAC and KEK B-factory experiments BaBar and Belle  provided much information  about standard model (SM), in particular in confirming SM predictions for CP violation based on Kobayashi-Maskawa (KM) mechanism~\cite{km}. Data from Tevatron confirmed many of the B-factory measurements. The LHCb experiment also started to provide interesting data about $B$ decays after the successful running of LHC.
The LHCb collaboration has recently reported evidences for CP asymmetries in $B^+$ decays into $\pi^+\pi^+\pi^-, \pi^+ K^+ K^-, K^+\pi^+\pi^-$, and $K^+ K^+ K^-$. We will refer to these decays as charged 3-body $B^+$ decays.

The CP asymmetries measured for the two $\Delta S = 1$, $K^+ \pi^+\pi^-$ and $K^+K^+K^-$  final states are~\cite{lhcb1}
\begin{eqnarray}
&&A_{CP}(K^+\pi^+\pi^-) = +0.032\pm 0.008(stat) \pm 0.004(syst)\pm 0.007(J/\psi K^+)\;,\nonumber\\
&&A_{CP}(K^+K^+K^-) = -0.043\pm 0.009(stat) \pm 0.003(syst)\pm 0.007(J/\psi K^+)\;,
\end{eqnarray}
which are $2.8\sigma$ and $3.7\sigma$ away from zero, respectively. Recently BaBar collaboration also reported their measurement~\cite{babar} of $A_{CP}(K^+K^+K^-) = -0.017^{+0.019}_{-0.014}\pm0.014$ which is consistent with the LHCb result within $1.1\sigma$.

The other two CP asymmetries are given by~\cite{lhcb2}
\begin{eqnarray}
&&A_{CP}(\pi^+\pi^+\pi^-) = +0.117\pm 0.021(stat) \pm 0.009(syst)\pm 0.007(J/\psi K^+)\;,\nonumber\\
&&A_{CP}(\pi^+K^+K^-) = -0.141\pm 0.040(stat) \pm 0.018(syst)\pm
0.007(J/\psi K^+)\;.
\end{eqnarray}
The significances are 4.2$\sigma$ and 3.0$\sigma$, respectively.

The branching ratios for these decays have also been measured with~\cite{brs}
\begin{eqnarray}
&&Br(\pi^+\pi^+\pi^-) = (15.2\pm 1.4)\times 10^{-6}\;,\;\;\;\;\;Br(\pi^+ K^+K^-) = (5.0\pm 0.7)\times 10^{-6}\;,\nonumber\\
&&Br(K^+K^+K^-) = (34.0\pm1.0)\times 10^{-6}\;,\;\;Br(K^+ \pi^+\pi^-) = (51.0\pm 3.0)\times 10^{-6}\;.
\end{eqnarray}

These charged 3-body $B^+$ decays can provide new information about the SM and for strong interaction which determine the hadronic matrix elements for $B$ decays. If flavor $SU(3)$ symmetry is a good symmetry for $B$ decays~\cite{su31} and the decay amplitude is dominated by the momentum independent contribution, it is expected that branching ratios for the pairs with $\Delta S=0$ and $\Delta S = 1$ have the following relations~\cite{gronau1},
\begin{eqnarray}
&&Br(\pi^+\pi^+\pi^-) = 2Br(\pi^+ K^+K^-)\;,\;\;\;\;\;\;Br(K^+K^+K^-) = 2Br(K^+\pi^+\pi^-)\;,\nonumber\\
&&A_{CP}(\pi^+\pi^+\pi^-) = A_{CP}(\pi^+ K^+K^-)\;,\;\;A_{CP}(K^+K^+K^-) = A_{CP}(K^+\pi^+\pi^-)\;. \label{pred1}
\end{eqnarray}

It is also expected that the rate asymmetries $\Delta_{CP} = \Gamma - \overline {\Gamma}$ to satisfy,
\begin{eqnarray}
\Delta_{CP}(\pi^+\pi^+\pi^-) = 2\Delta_{CP}(\pi^+ K^+K^-) = -
\Delta_{CP}(K^+K^+K^-) = -2\Delta_{CP}(K^+\pi^+\pi^-)\;.
\end{eqnarray}
 Note that $\Delta_{CP}$ and $A_{CP}$ is related by $A_{CP} = \tau_B \Delta_{CP}/2Br$. Therefore one can obtain relations between asymmetries $A_{CP}$ for different decays.

Inspection of data from the LHCb, one finds that only $\Delta_{CP}(\pi^+\pi^+\pi^-) = -\Delta_{CP}(K^+K^+K^-)$ is in agreement with data, the other predictions described above do not hold, in particular the relative signs are different from $SU(3)$ predictions between  $A_{CP}(\pi^+\pi^+\pi^-)$ and $A_{CP}(\pi^+ K^+K^-)$,  and $A_{CP}(K^+K^+K^-)$ and $A_{CP}(K^+\pi^+\pi^-)$ pairs.  If the experimental data are further confirmed, one needs to go beyond the $SU(3)$ conserving momentum independent contribution to the decay amplitudes to see if the decay pattern observed can be explained. One of the possible sources causing the deviation may come from $SU(3)$ breaking effects due to light quark difference of $m_{u}$, $m_d$ and $m_s$.

The flavor $SU(3)$ symmetry is expected to be only an approximate symmetry because $u$, $d$ and $s$ quarks have different masses. The $SU(3)$ breaking effect is at level of 20 percent for the $\pi$ and $K$ decay constants $f_\pi$ and $f_K$. For 2-body pseudoscalar octet meson $B$ decays, the flavor $SU(3)$ symmetry works reasonably well although there are some breakings~\cite{su32} and some predictions work very well, such as rate differences between some of the $\Delta S=0$ and $\Delta S=1$ two-body pseudoscalar meson $B$ decays~\cite{he1,he2}. It is natural to ask that whether with $SU(3)$ breaking effects, one can explain the above mentioned charged 3-body $B^+$ decay pattern. The $SU(3)$ breaking quark mass contributions can be considered as sub-leading corrections to the leading $SU(3)$ breaking effects mentioned earlier.

Inclusion of contributions from quark masses should be considered consistently with other possible sub-leading contributions. To this end we note that the $K$ and $\pi$ masses squared are proportional to light quark masses in chiral perturbation theory, to include light quark mass contribution consistently one should also consider terms with two derivatives. By doing this it will allow new terms which do not exist in the momentum indpendent $SU(3)$ amplitude, such as terms like, $ (\partial^\mu K^+ \pi^+ - \partial^\mu \pi^+ K^+)\partial_\mu K^- $ and $(\partial^\mu K^+ \pi^+ - \partial^\mu \pi^+ K^+)\partial_\mu \pi^- $. These terms will contribute to $B^+ \to \pi^+ K^+ K^-,\;\; K^+\pi^+\pi^-$. There is no equivalent contribution to $B^+ \to \pi^+\pi^+\pi^-$ and $B^+ \to K^+ K^+ K^-$. Therefore such contributions can lead to deviation from Eq.(\ref{pred1}) and may help to explain data.

In this work we study  the above mentioned two types of effects in these decays in the framework of flavor $SU(3)$ symmetry. We find that to explain the observed decay pattern large $SU(3)$ breaking effects are needed.

\section{$SU(3)$ conserving momentum independent amplitudes}

We start with the description of $B$ decays into three pseudoscalar octet mesons from flavor $SU(3)$ symmetry.
The leading quark level effective Hamiltonian up to one loop level in
electroweak interaction for hadronic charmless $B$ decays in the SM can be written as
\begin{eqnarray}
 H_{eff}^q = {4 G_{F} \over \sqrt{2}} [V_{ub}V^{*}_{uq} (c_1 O_1 +c_2 O_2)
   - \sum_{i=3}^{12}(V_{ub}V^{*}_{uq}c_{i}^{uc} +V_{tb}V_{tq}^*
   c_i^{tc})O_{i}],
\end{eqnarray}
where $q$ can be $d$ or $s$the coefficients
$c_{1,2}$ and $c_i^{jk}=c_i^j-c_i^k$, with $j$ and $k$ indicate the internal quark,
are the Wilson Coefficients (WC).  The tree WCs are of order one with, $c_1=-0.31$, and $c_2 = 1.15$. The penguin
WCs are much smaller with the largest one $c_6$ to be $-0.05$. These
WC's have been evaluated by several groups~\cite{heff}. $V_{ij}$ are the KM matrix elements.
In the above the factor $V_{cb}V_{cq}^*$ has
been eliminated using the unitarity property of the KM matrix.

The operators $O_i$ are given by
\begin{eqnarray}
\begin{array}{ll}
O_1=(\bar q_i u_j)_{V-A}(\bar u_i b_j)_{V-A}\;, &
O_2=(\bar q u)_{V-A}(\bar u b)_{V-A}\;,\\
O_{3,5}=(\bar q b)_{V-A} \sum _{q'} (\bar q' q')_{V \mp A}\;,&
O_{4,6}=(\bar q_i b_j)_{V-A} \sum _{q'} (\bar q'_j q'_i)_{V \mp A}\;,\\
O_{7,9}={ 3 \over 2} (\bar q b)_{V-A} \sum _{q'} e_{q'} (\bar q' q')_{V \pm A}\;,\hspace{0.3in} &
O_{8,10}={ 3 \over 2} (\bar q_i b_j)_{V-A} \sum _{q'} e_{q'} (\bar q'_j q'_i)_{V \pm A}\;,\\
O_{11}={g_s\over 16\pi^2}\bar q \sigma_{\mu\nu} G^{\mu\nu} (1+\gamma_5)b\;,&
O_{12}={Q_b e\over 16\pi^2}\bar q \sigma_{\mu\nu} F^{\mu\nu} (1+\gamma_5)b.
\end{array}
\end{eqnarray}
where $(\bar a b)_{V-A} = \bar a \gamma_\mu (1-\gamma_5) b$, $G^{\mu\nu}$ and
$F^{\mu\nu}$ are the field strengths of the gluon and photon, respectively.

At the hadron level, the decay amplitude can be generically written as
\begin{eqnarray}
A = \langle final\;state|H_{eff}^q|\bar {B}\rangle = V_{ub}V^*_{uq} T(q) + V_{tb}V^*_{tq}P(q)\;,
\end{eqnarray}
where $T(q)$ contains contributions from the $tree$ as well as $penguin$ due to charm and up
quark loop corrections to the matrix elements,
while $P(q)$ contains contributions purely from
one loop $penguin$ contributions. $B$ indicates one of the $B^+$, ${B}^0_d$ and ${B}^0_s$. $B_i = (B^+,  B^0,  B^0_s)$ forms an $SU(3)$ triplet.

The flavor $SU(3)$ symmetry transformation properties for operators $O_{1,2}$, $O_{3-6, 11,12}$, and $O_{7-10}$ are: $\bar 3_a + \bar 3_b +6 + \overline {15}$,
$\bar 3$, and $\bar 3_a + \bar 3_b +6 + \overline {15}$, respectively.
We indicate these representations by matrices in $SU(3)$ flavor space by $H(\bar 3)$, $H(6)$ and $H(\overline{15})$.
For $q=d$, the non-zero entries of the matrices $H(i)$ are given by~\cite{he1}
\begin{eqnarray}
H(\bar 3)^2 &=& 1\;,\;\;
H(6)^{12}_1 = H(6)^{23}_3 = 1\;,\;\;H(6)^{21}_1 = H(6)^{32}_3 =
-1\;,\nonumber\\
H(\overline {15} )^{12}_1 &=& H(\overline {15} )^{21}_1 = 3\;,\; H(\overline
{15} )^{22}_2 =
-2\;,\;
H(\overline {15} )^{32}_3 = H(\overline {15} )^{23}_3 = -1\;.
\end{eqnarray}
And for $q = s$, the non-zero entries are
\begin{eqnarray}
H(\bar 3)^3 &=& 1\;,\;\;
H(6)^{13}_1 = H(6)^{32}_2 = 1\;,\;\;H(6)^{31}_1 = H(6)^{23}_2 =
-1\;,\nonumber\\
H(\overline {15} )^{13}_1 &=& H(\overline {15} ) ^{31}_1 = 3\;,\; H(\overline
{15} )^{33}_3 =
-2\;,\;
H(\overline {15} )^{32}_2 = H(\overline {15} )^{23}_2 = -1\;.
\end{eqnarray}

These properties enable one to
write the decay amplitudes for $B \to PPP$ decays in only a few $SU(3)$ invariant amplitudes~\cite{su31}. Here $P$ is one of the mesons in the pseudoscalar octet meson $M=(M_{ij})$ which is given by,
\begin{eqnarray}
M= \left ( \begin{array}{ccc}
{\pi^0\over \sqrt{2}} + {\eta_8\over \sqrt{6}}&\pi^+&K^+\\
\pi^-&-{\pi^0\over \sqrt{2}} + {\eta_8\over \sqrt{6}}& K^0\\
K^-&\bar K^0&-{2\eta_8\over \sqrt{6}}
\end{array} \right ).
\end{eqnarray}

Construction of $B^+ \to PPP$ decay amplitude can be done order by order by using three $M$'s, $B$, and the Hamiltonian $H$, and also derivatives on the mesons to form $SU(3)$. Let us discuss $SU(3)$ conserving
momentum independent amplitudes in the following.

For the $T(q)$ amplitude, we have
\begin{eqnarray}
T(q)&&=a^T(\overline{3})B_{i}H^{i}(\overline{3})M^{j}_{k}M^{k}_{l}M^{l}_{j}+b^T(\overline{3})H^{i}(\overline{3})M^{j}_{i}B_{j}M^{k}_{l}M^{l}_{k} +c^T(\overline{3})H^{i}(\overline{3})M^{l}_{i}M^{j}_{l}M^{k}_{j}B_{k}\nonumber\\
&&+a^T(6)B_{i}H^{ij}_{k}(6)M^{k}_{j}M^{l}_{n}M^{n}_{l}+b^T(6)B_{i}H^{ij}_{k}(6)M^{k}_{l}M^{l}_{n}M^{n}_{j}\nonumber\\
&&+c^T(6)B_{i}H^{jk}_{l}(6)M^{i}_{j}M^{n}_{k}M^{l}_{n}+d^T(6)B_{i}H^{jk}_{l}(6)M^{i}_{n}M^{l}_{j}M^{n}_{k}\nonumber\\
&&+a^T(\overline {15})B_{i}H^{ij}_{k}(\overline{15})M^{k}_{j}M^{l}_{n}M^{n}_{l}+b^T(\overline{15})B_{i}H^{ij}_{k}(\overline{15})M^{k}_{l}M^{l}_{n}M^{n}_{j}\nonumber\\
&&+c^T(\overline{15})B_{i}H^{jk}_{l}(\overline{15})M^{i}_{j}M^{n}_{k}M^{l}_{n}+d^T(\overline{15})B_{i}H^{jk}_{l}(\overline{15})M^{i}_{n}M^{l}_{j}M^{n}_{k}\;.\label{su3s}
\end{eqnarray}
One can write similar amplitude $P(q)$ for the penguin contributions.

The coefficients $a(i)$, $b(i)$, $c(i)$ and $d(i)$ are constants which contain the WCs and information about QCD dynamics.
Several groups have studied $B\to PPP$ decays~~\cite{he3,fajfer,hycheng,london}.
In general there may be resonant contributions~\cite{he3,fajfer,hycheng,london,yadong} due to exchange of intermediate particles resulting in the parameters $a(i)$ to $d(i)$ to be dependent of momenta exchanged, such as dependent on the s, t and u variables.
There may be other contributions which can also make the decay amplitudes momentum dependent.
The LHCb has also measured CP asymmetries for localized regions of phase space according to invariant masses of $\pi^+\pi^-$ and $K^+K^-$ with larger asymmetries~\cite{lhcb1,lhcb2}. If confirmed, this indicates that the decay amplitudes have momentum dependent from exchanging particles~\cite{yadong}. If only considering the momentum independent contributions given above, one will not be able to estimate the localized CP asymmetries. We will not consider localized CP asymmetries, but concentrate on CP violation in these decays with phase space integrated over.


Expanding the above $T(q)$ amplitude, one can extract the decay amplitudes
$T(PPP)$ for $B^{+} \to
\pi^{+}\pi^{+}\pi^{-}$, $B^{+} \to \pi^+ K^{+}K^{-}$,  and $B^{+} \to
K^{+}\pi^{+}\pi^{-}$,   $ B^{+} \to K^{+}K^{+}K^{-}$. We find that they are all equal. Indicating it by $T$, we have
\begin{eqnarray}
T&&=2b^T(\overline{3})+c^T(\overline{3})+2a^T(6)+b^T(6)-c^T(6)-d^T(6)\nonumber\\&&+6a^T(\overline{15})+3b(\overline{15})+c^T(\overline{15})+3d^T(\overline{15})\;. \label{leading}
\end{eqnarray}
Similarly, we find the same situation for the penguin amplitude $P$. $P$ amplitude can be obtained by replacing $T$ by $P$ in the above expression.

The amplitudes for the four charged 3-body $B^{+}$ decays can be written as
\begin{eqnarray}
&&A(B^{+} \to
\pi^{+}
\pi^{+}\pi^{-})=V_{ub}^{*}V_{ud}T+V_{tb}^{*}V_{td}P\;,\nonumber
\\
&&A(B^{+} \to
\pi^{+} K^{+}K^{-})=V_{ub}^{*}V_{ud}T+V_{tb}^{*}V_{td}P\;,\nonumber
\\
&&A(B^{+} \to
K^{+} \pi^{+}\pi^{-})=V_{ub}^{*}V_{us}T+V_{tb}^{*}V_{ts}P\;,\nonumber
\\
&&A(B^{+} \to K^{+}
K^{+}K^{-})=V_{ub}^{*}V_{us}T+V_{tb}^{*}V_{ts}P\;.
\end{eqnarray}
For amplitude involving identical particles, we use the convention to include the identical particle factor at the branching ratio calculation stage.

The corresponding decay amplitudes for $B^-$ decays can be obtained from the above by replacing $V_{ub}V_{uq}^*$ and $V_{tb}V_{tq}^*$ by
$V_{ub}^*V_{uq}$ and $V_{tb}^*V_{tq}$, respectively.

In the $SU(3)$ limit, we have
\begin{eqnarray}
Br(\pi^+\pi^+\pi^-) = 2Br(\pi^+ K^+K^-)\;,\;\;\;\;Br(K^+\pi^+\pi^-) = 2Br(K^+ K^+K^-)\;. \label{another1}
\end{eqnarray}
The factor of 2 in the above equations are due to identical particles $\pi^+\pi^+$ and $K^+K^+$ in $B^+\to \pi^+\pi^+\pi^-$ and $B^+\to K^+K^+K^-$.

Because the KM factors involved for the above two $\Delta S = 0$ and two $\Delta S =1$ decays are different, their branching ratios are not expected to be the same. However, because the relation~\cite{jarlskog} of KM
matrix element
$Im(V_{ub}V_{ud}^{*}V_{tb}^{*}V_{td})=-Im(V_{ub}V_{us}^{*}V_{tb}^{*}V_{ts})$,
the CP violating rate difference of the $\Delta S =0$ and $\Delta S = 1$ are related.
We have
\begin{eqnarray}
&&2\Delta(K^{+} \pi^{+}\pi^{-})=\Delta(K^{+}
K^{+}K^{-})=-\Delta(\pi^{+} \pi^{+}\pi^{-})=-2\Delta(\pi^{+}
K^{+}K^{-})\;,
\end{eqnarray}
which leads to the relations for $A_{CP}$ given by
 \begin{eqnarray}
&&{A_{CP}(\pi^{+}K^{+}K^{-}) \over A_{CP}(K^{+}
 \pi^{+}\pi^{-})}=-{Br(K^{+}
 \pi^{+}\pi^{-}) \over Br(\pi^{+}K^{+}K^{-})}\;,\;\;{A_{CP} (\pi^{+} \pi^{+}\pi^{-}) \over A_{CP} (K^{+}
 K^{+}K^{-})}=-{Br(K^{+} K^{+}K^{-}) \over Br(\pi^{+}
 \pi^{+}\pi^{-})}\;,\nonumber\\
 &&{A_{CP}(\pi^{+}K^{+}K^{-}) \over A_{CP}(K^{+} K^{+}K^{-}) }=-{Br(K^{+} K^{+}K^{-})\over Br(\pi^{+}K^{+}K^{-})}\;,\;\;{A_{CP}(\pi^{+} \pi^{+}\pi^{-}) \over A_{CP}(K^{+}
 \pi^{+}\pi^{-})}=-{Br(K^{+} \pi^{+}\pi^{-})\over Br(\pi^{+}
 \pi^{+}\pi^{-})}\;. \label{cpa}
\end{eqnarray}

The LHCb data obviously do not support the branching ratio relations given above. The relations for CP asymmetry $A_{CP}$
do not agree with data either, except the ratio $A_{CP} (\pi^{+} \pi^{+}\pi^{-}) / A_{CP} (K^{+}
 K^{+}K^{-})$. The LHCb data $A_{CP} (\pi^{+} \pi^{+}\pi^{-}) / A_{CP} (K^{+}
 K^{+}K^{-}) = -2.7\pm 0.9$ agrees with the predicted value~\cite{gronau1} $-2.2\pm 0.2$ very well using Eq.(\ref{cpa}).
If experimental data at the LHCb will be further confirmed, one needs to include contributions from beyond the  $SU(3)$ conserving momentum independent effects to explain the data. There may be different sources which can cause the deviations, one of the possibilities is
the $SU(3)$ breaking effects. To be consistent in carrying out the analysis, as mentioned earlier, one also needs to take into account contributions from terms with two derivatives. In the next section we study these contributions.

\begin{table}[h!]
\begin{centering}
\begin{tabular}{|c|c|c|}
\hline  ${A_{CP}(\Delta S=0) / A_{CP}(\Delta S =1)}$&Momentum independent & Data\\ 
&amplitude predictions&\\
\hline ${A_{CP}(\pi^{+}K^{+}K^{-}) / A_{CP}(K^{+}
 \pi^{+}\pi^{-})}$ & $-10.2 \pm 1.5$ & $-4.4 \pm 2.0$\\
 \hline $A_{CP} (\pi^{+} \pi^{+}\pi^{-}) / A_{CP} (K^{+}
 K^{+}K^{-})$ & $-2.2 \pm0.2$ & $-2.7 \pm 0.9$\\
 \hline $A_{CP}(\pi^{+}K^{+}K^{-}) / A_{CP}(K^{+} K^{+}K^{-})$ & $-6.8 \pm
 1.1$ & $+3.3 \pm 1.4$\\
 \hline $A_{CP}(\pi^{+} \pi^{+}\pi^{-}) / A_{CP}(K^{+}
 \pi^{+}\pi^{-})$ & $-3.4 \pm 0.3$ & $+3.7 \pm 1.5$ \\
 \hline
\end{tabular}
\caption{Comparison of $SU(3)$ conserving momentum independent amplitude predictions and data for $A_{CP}(\Delta S=0)/A_{CP}(\Delta S =1)$.}
\end{centering}\end{table}

\section{Contributions from $SU(3)$ breaking and derivative terms}

\subsection{$SU(3)$ Breaking Due To Light Quark Masses}

Flavor $SU(3)$ symmetry breaking effects come from difference in
masses of $u$, $d$ and $s$ quarks. Under $SU(3)$, the mass matrix
can be viewed as combinations of representations from $3\times \bar
3$, to matching the ($u$, $d$, $s$) transformation property as a
fundamental representation, which contains an 1 and an 8 irreducible
representations. The diagonalized  mass matrix can be expressed as a
linear combination of the identity matrix $I$, and the Gell-Mann
matrices $\lambda_3$ and $\lambda_8$. Compared with $s$-quark mass
$m_s$, the $u$ and $d$ quark masses $m_{u,d}$  are much smaller
which can be neglected and therefore the term proportional to
$\lambda_3$ disappears. The s-quark mass is the main source for
flavor $SU(3)$ symmetry breaking which is of a diaganol matrix form,
$m_s W$ with $W=dig(0,0,1)$.
It can be further decomposed into  $I$ and $\lambda_8$. The identity $I$ part contributes to the $B$ decay amplitudes in a similar way as that given in eq. (\ref{su3s}) which can be absorbed into the coefficients $a(i)$ to $d(i)$. Only $\lambda_8$ piece will contribute to the $SU(3)$ breaking effects. We will use this to construct $SU(3)$ breaking decay amplitudes and indicate it by~\cite{gronau2}.
\begin{eqnarray}
W = (W^i_j) =\left ( \begin{array}{ccc}
 1 & 0 & 0\\
 0 & 1& 0 \\
0 &  0 & -2
\end{array}
\right )\;.
\end{eqnarray}

To construct relevant decay amplitudes for charged 3-body $B^+$ decays, one first breaks the contraction of indices at any joint in eq. (\ref{su3s}), and inserts a W in between,  and then contracts all indices appropriately as shown in Appendix A. Each possible way will be associated with a coefficient which we will treat as a free parameter.

Expanding the above expression, we obtain the $\Delta T$ amplitudes as follows.
\begin{eqnarray}
&&\Delta T(\pi^{+}\pi^{+}\pi^{-})=2b^T_{1}(\overline{3})+2b^T_{2}(\overline{3})+2b^T_{3}(\overline{3})+c^T_{1}(\overline{3})+c^T_{2}(\overline{3})+c^T_{3}(\overline{3})+c^T_{4}(\overline{3})\nonumber\\
&&\qquad\qquad\qquad+2a^T_{1}(6)+2a^T_{2}(6)+2a^T_{3}(6)+2a^T_{4}(6)+b^T_{1}(6)+b^T_{2}(6)+b^T_{3}(6) \nonumber\\
&&\qquad\qquad\qquad+b^T_{4}(6)+b^T_{5}(6)-c^T_{1}(6)-c^T_{2}(6)-c^T_{3}(6)-c^T_{4}(6)-c^T_{5}(6) \nonumber\\
&&\qquad\qquad\qquad-d^T_{1}(6)-d^T_{2}(6)-d^T_{3}(6)-d^T_{4}(6)-d^T_{5}(6)+6a^T_{1}(\overline{15}) \nonumber\\
&&\qquad\qquad\qquad+6a^T_{2}(\overline{15})+6a^T_{3}(\overline{15})+6a^T_{4}(\overline{15})+3b^T_{1}(\overline{15})+3b^T_{2}(\overline{15})+3b^T_{3}(\overline{15})+3b^T_{4}(\overline{15})\nonumber\\
&&\qquad\qquad\qquad+3b^T_{5}(\overline{15})+c^T_{1}(\overline{15})+c^T_{2}(\overline{15})+c^T_{3}(\overline{15})+c^T_{4}(\overline{15})+c^T_{5}(\overline{15})+3d^T_{1}(\overline{15}) \nonumber\\
&&\qquad\qquad\qquad+3d^T_{2}(\overline{15})+3d^T_{3}(\overline{15})+3d^T_{4}(\overline{15})+3d^T_{5}(\overline{15})\;,    \nonumber\\
&&\Delta T(K^{+}K^{-}\pi^{+})=2b^T_{1}(\overline{3})+2b^T_{2}(\overline{3})-b^T_{3}(\overline{3})+c^T_{1}(\overline{3})+c^T_{2}(\overline{3})+c^T_{3}(\overline{3})-2c^T_{4}(\overline{3})\nonumber\\
&&\qquad\qquad\qquad+2a^T_{1}(6)+2a^T_{2}(6)+2a^T_{3}(6)-a^T_{4}(6)+b^T_{1}(6)+b^T_{2}(6)+b^T_{3}(6)\nonumber\\
&&\qquad\qquad\qquad-2b^T_{4}(6)+b^T_{5}(6)-c^T_{1}(6)+2c^T_{2}(6)-4c^T_{3}(6)-c^T_{4}(6)+2c^T_{5}(6)\nonumber\\
&&\qquad\qquad\qquad-d^T_{1}(6)-d^T_{2}(6)-d^T_{3}(6)-d^T_{4}(6)+2d^T_{5}(6)+6a^T_{1}(\overline{15})\nonumber\\
&&\qquad\qquad\qquad+6a^T_{2}(\overline{15})+6a^T_{3}(\overline{15})-3a^T_{4}(\overline{15})+3b^T_{1}(\overline{15})+3b^T_{2}(\overline{15})+3b^T_{3}(\overline{15})-6b^T_{4}(\overline{15})\nonumber\\
&&\qquad\qquad\qquad+3b^T_{5}(\overline{15})+c^T_{1}(\overline{15})+4c^T_{2}(\overline{15})+4c^T_{3}(\overline{15})+7c^T_{4}(\overline{15})-8c^T_{5}(\overline{15})+3d^T_{1}(\overline{15})\nonumber\\
&&\qquad\qquad\qquad+3d^T_{2}(\overline{15})+3d^T_{3}(\overline{15})+3d^T_{4}(\overline{15})-6d^T_{5}(\overline{15})\;,     \nonumber\\
&&\Delta T(K^{+}\pi^{+}\pi^{-})=-4b^T_{1}(\overline{3})+2b^T_{2}(\overline{3})+2b^T_{3}(\overline{3})-2c^T_{1}(\overline{3})+c^T_{2}(\overline{3})+c^T_{3}(\overline{3})+c^T_{4}(\overline{3})\nonumber\\
&&\qquad\qquad\qquad+2a^T_{1}(6)-4a^T_{2}(6)+2a^T_{3}(6)+2a^T_{4}(6)+b^T_{1}(6)-2b^T_{2}(6)+b^T_{3}(6)\nonumber\\
&&\qquad\qquad\qquad+b^T_{4}(6)+b^T_{5}(6)-c^T_{1}(6)-c^T_{2}(6)+2c^T_{3}(6)-c^T_{4}(6)-c^T_{5}(6)\nonumber\\
&&\qquad\qquad\qquad-d^T_{1}(6)+2d^T_{2}(6)-d^T_{3}(6)-d^T_{4}(6)-d^T_{5}(6)+6a^T_{1}(\overline{15})\nonumber\\
&&\qquad\qquad\qquad-12a^T_{2}(\overline{15})+6a^T_{3}(\overline{15})+6a^T_{4}(\overline{15})+3b^T_{1}(\overline{15})-6b^T_{2}(\overline{15})+3b^T_{3}(\overline{15})+3b^T_{4}(\overline{15})\nonumber\\
&&\qquad\qquad\qquad+3b^T_{5}(\overline{15})+c^T_{1}(\overline{15})-5c^T_{2}(\overline{15})+4c^T_{3}(\overline{15})+c^T_{4}(\overline{15})+c^T_{5}(\overline{15})+3d^T_{1}(\overline{15})\nonumber\\
&&\qquad\qquad\qquad-6d^T_{2}(\overline{15})+3d^T_{3}(\overline{15})+3d^T_{4}(\overline{15})+3d^T_{5}(\overline{15})\;,     \nonumber\\
&&\Delta T(K^{+}K^{-}K^{+})=-4b^T_{1}(\overline{3})+2b^T_{2}(\overline{3})-2b^T_{3}(\overline{3})-2c^T_{1}(\overline{3})+c^T_{2}(\overline{3})+c^T_{3}(\overline{3})-2c^T_{4}(\overline{3})\nonumber\\
&&\qquad\qquad\qquad+2a^T_{1}(6)-4a^T_{2}(6)+2a^T_{3}(6)-a^T_{4}(6)+b^T_{1}(6)-2b^T_{2}(6)+b^T_{3}(6)\nonumber\\
&&\qquad\qquad\qquad-2b^T_{4}(6)+b^T_{5}(6)-c^T_{1}(6)+2c^T_{2}(6)-c^T_{3}(6)-c^T_{4}(6)+2c^T_{5}(6)\nonumber\\
&&\qquad\qquad\qquad-d^T_{1}(6)+2d^T_{2}(6)-d^T_{3}(6)-d^T_{4}(6)+2d^T_{5}(6)+6a^T_{1}(\overline{15})\nonumber\\
&&\qquad\qquad\qquad-12a^T_{2}(\overline{15})+6a^T_{3}(\overline{15})-3a^T_{4}(\overline{15})+3b^T_{1}(\overline{15})-6b^T_{2}(\overline{15})+3b^T_{3}(\overline{15})-6b^T_{4}(\overline{15})\nonumber\\
&&\qquad\qquad\qquad+3b^T_{5}(\overline{15})+c^T_{1}(\overline{15})-2c^T_{2}(\overline{15})+7c^T_{3}(\overline{15})+7c^T_{4}(\overline{15})-8c^T_{5}(\overline{15})+3d^T_{1}(\overline{15})\nonumber\\
&&\qquad\qquad\qquad-6d^T_{2}(\overline{15})+3d^T_{3}(\overline{15})+3d^T_{4}(\overline{15})-6d^T_{5}(\overline{15})\;.
\end{eqnarray}

Had one used $W = diag(0,0,1)$ for $W$, the correction to $\Delta T(\pi^{+}\pi^{+}\pi^{-})$ would vanish. But both ways of calculating the corrections are equivalent. This can be easily understood by substracting the same among correction as to $\Delta T(\pi^{+}\pi^{+}\pi^{-})$ from the other three amplitudes and absorb them into the leading order amplitudes in Eq.(\ref{leading}).


Note that the amplitudes are not all independent, but satisfy,
\begin{eqnarray}
\Delta T(K^{+}K^{-}K^{+})-\Delta T(K^{+}\pi^{+}\pi^{-})=\Delta T(K^{+}K^{-}\pi^{+})-\Delta T(\pi^{+}\pi^{+}\pi^{-})\;.
\end{eqnarray}
Similarly one can obtain the penguin amplitudes $\Delta P$.

\subsection{Terms With Derivatives}

The lowest order terms with derivatives lead to two powers of momentum dependent.
One can obtain relevant terms by taking two times of derivatives on each of the terms in Eq.(\ref{su3s}) and  then collectting them together. Not all of the terms are independent. If the two derivatives are taken on one field, $\partial^2B$ or $\partial^2 M$, by using on-shell conditions of the particles, these terms will be proportional to the ones already existed which can be absorbed into redefinition of the coefficients in Eq.(\ref{su3s}) in the $SU(3)$ limit. When $SU(3)$ is broken due to quark mass differences, terms containing $\partial^2 K = m^2_K K$ and $\partial^2\pi = m^2_\pi \pi$ will generate $SU(3)$ breaking terms. However, they will not create new terms already discussed in the previous section. We find that independent terms can only come from taking derivatives on two different fields. For example after taking derivatives for $B_i H^i(\overline 3) M^j_k M^k_l M^l_j$, we have the following terms
\begin{eqnarray}
&&(\partial_\mu B_i) H^i(\overline 3) (\partial^\mu M^j_k) M^k_l M^l_j,\;\;(\partial_\mu B_i) H^i(\overline 3) M^j_k (\partial^\mu M^k_l) M^l_j,\;\;(\partial_\mu B_i) H^i(\overline 3) M^j_k M^k_l (\partial^\mu M^l_j)\;,\nonumber\\
&&B_i H^i(\overline 3) (\partial_\mu M^j_k) (\partial^\mu M^k_l) M^l_j,\;\;B_i H^i(\overline 3) (\partial_\mu M^j_k ) M^k_l (\partial^\mu M^l_j),\;\;B_i H^i(\overline 3) M^j_k (\partial_\mu M^k_l) (\partial^\mu M^l_j)\;.\nonumber\\
\end{eqnarray}

In the above the lower and upper indices are contracted. One may wonder whether the indices which are contracted by $\epsilon^{ijk}$ and $\epsilon_{ijk}$ give new terms. This is not the case because of the identity
\begin{eqnarray}
\epsilon_{ijk}\epsilon^{abc} = \left | \begin{array}{ccc}
 \delta^a_i & \delta^b_i & \delta^c_i\\
 \delta^a_j & \delta^b_j & \delta^c_j\\
\delta^a_k & \delta^b_k & \delta^c_k\\
\end{array}
\right |\;.
\end{eqnarray}
The full list of the possible terms are given in Appendix B.

Expanding terms in Appendix B, one obtains the lowest derivative amplitude $T^p$ for $B^+$ decay into three charged PPP mesons in six different forms depending on how the derivatives are taken. For $\Delta S = -1$ case. We have
\begin{eqnarray}
&&(a)\;(\partial_\mu B^+) (\partial^\mu K^+) [K^+ K^- + \pi^+\pi^-],
\;\;\;\;\;\;\;\;\;(b)\;(\partial_\mu B^+)  K^+ [K^+ (\partial^\mu K^-) + \pi^+(\partial^\mu \pi^-)], \nonumber\\
&&(c)\;(\partial_\mu B^+) K^+ [(\partial^\mu K^+) K^- + (\partial^\mu \pi^+)\pi^-],
\;\;\;(d)\;B^+ K^+ [(\partial_\mu K^+) (\partial^\mu K^-) + (\partial_\mu \pi^+)(\partial^\mu \pi^-)], \nonumber\\
&&(e)\;B^+ \partial_\mu K^+ [K^+ (\partial^\mu K^-) + \pi^+(\partial^\mu \pi^-)], \;\;\;\;\;(f)\;B^+ \partial_\mu K^+ [(\partial^\mu K^+) K^- + (\partial^\mu \pi^+)\pi^-].
\end{eqnarray}
In the above, $B^+$ is going in and the three light pesudoscalar mesons are going out.

To see how these terms change the leading $SU(3)$ amplitude, it is convenient to use the following six different terms to exppress the above
\begin{eqnarray}
&&(1)\;\;(\partial_\mu B^+) [(\partial^\mu K^+) \pi^+ - K^+(\partial^\mu \pi^+)]\pi^-\;,
\;\;(2)\;\;(\partial_\mu B^+)  K^+ [K^+ (\partial^\mu K^-) +\pi^+(\partial^\mu \pi^-)]\;, \nonumber\\
&&(3)\;\;(\partial_\mu B^+) [K^+ (\partial^\mu K^+) K^- + {1\over 2}((\partial^\mu (K^+)\pi^+ + K^+ (\partial^\mu \pi^+))\pi^-) ]\;,\nonumber\\
&&(4)\;\;B^+ [(\partial^\mu K^+) \pi^+ - K^+(\partial^\mu \pi^+)]\partial_\mu \pi^-\;\;\;,\;\;(5)\;\;B^+ \partial_\mu K^+ [(\partial^\mu K^+) K^- + (\partial^\mu \pi^+)\pi^-]\;, \nonumber\\
&&(6)\;\;B^+ [(\partial_\mu K^+)K^+ (\partial^\mu K^-) +{1\over 2}((\partial_\mu K^+) \pi^+ + K^+ (\partial_\mu \pi^+)) \partial^\mu\pi^-)]
\;. \label{p-term}
\end{eqnarray}
The terms (1), (4), ((3) and (6)) are obtained by $(a) - (c))$
($((a)+(c))/2$), and by  $(e) - (d)$ ($((e)+(d))/2$), respectively.

There are no analogous terms in the $SU(3)$ symmetric momentum independent contribution for the first two terms above. The existence of these terms make the amplitude
for $B^+ \to K^+K^+ K^-$ different from that for $B^+ \to K^+ \pi^+\pi^-$, and therefore providing another source of violating the relation in Eq.(\ref{another1}). Similarly, one can work out the independent terms for $\Delta S = 0$ decays.
Note that these terms contribute to the zero U-spin amplitude $A_0$ formed by the two charged mesons in the final state in the U-spin analysis\cite{gronau-u,xlh-u}. One should, however, be careful that with just the momentum independent amplitudes, $A_0$ is zero, due to Bose-Einstein statistics. It is not correct to make $A_0$ momentum independent\cite{xlh-u}. Direct U-spin construction of decay amplitudes obtain the same structure of momentum dependence to the same order as the above\cite{xlh-u}.

Expanding the terms in Appendix B, the $\Delta S = -1$ amplitudes $T^p$ are proportional to
\begin{eqnarray}
{1\over m_B^2}\left (\alpha_{1}(1)+\alpha_{2})2)+\alpha_{3})3)+
\alpha_{4}(4)+\alpha_{5}(5)+\alpha_{6} (6)\right )\;.
\end{eqnarray}
In the above, we have normalized the dimension of the coefficients $\alpha_i$ so that they are dimensionless.
Similarly, one can define the amplitude $P^p$ for the penguin contribution. The coefficients  $\alpha_i$ are given in terms of the coefficients in Appendix B. A similar expressions apply to the $\Delta S = 0$ amplitudes.

Replacing $\partial^\mu$ by momentum $p^\mu$ in the above expressions, we obtain the tree momentum dependent
amplitude $T^p$
\begin{eqnarray}
&&T^p(K^{+}(p_{1})K^{+}(p_{2})K^{-}(p_{3}))\nonumber\\
&&={1\over 2 m_B^2}\left (2\alpha_{2}p_{B}\cdot
p_{3}+\alpha_{3}p_{B}\cdot(p_{1}+p_{2})+2\alpha_{5}p_{1}\cdot
p_{2}+\alpha_{6}(p_{1}+p_{2})\cdot p_{3}\right )\;,\nonumber\\
&&T^p(K^{+}(p_{1})\pi^{+}(p_{2})\pi^{-}(p_{3}))\nonumber\\
&&={1\over 2 m_B^2}\left (
2\alpha_{2}p_{B}\cdot
p_{3}+\alpha_{3}p_{B}\cdot(p_{1}+p_{2})+2\alpha_{5}p_{1}\cdot
p_{2}+\alpha_{6}(p_{1}+p_{2})\cdot p_{3}\right . \nonumber\\
&&\left .+2(\alpha_{1}p_{B}\cdot (p_{1} -p_{2}) + \alpha_{4}(p_{1}-p_2)\cdot
p_{3} )\right )\;,\\
&&T^p(\pi^{+}(p_{1})\pi^{+}(p_{2})\pi^{-}(p_{3}))\nonumber\\
&&={1\over 2 m_B^2}\left (2\alpha_{2}p_{B}\cdot
p_{3}+\alpha_{3}p_{B}\cdot(p_{1}+p_{2})+2\alpha_{5}p_{1}\cdot
p_{2}+\alpha_{6}(p_{1}+p_{2})\cdot p_{3}\right )\;,\nonumber\\
&&T^p(\pi^{+}(p_{1})K^{+}(p_{2})K^{-}(p_{3}))\nonumber\\
&&={1\over 2 m_B^2}\left (
2\alpha_{2}p_{B}\cdot
p_{3}+\alpha_{3}p_{B}\cdot(p_{1}+p_{2})+2\alpha_{5}p_{1}\cdot
p_{2}+\alpha_{6}(p_{1}+p_{2})\cdot p_{3}\right . \nonumber\\
&&\left .+2(\alpha_{1}p_{B}\cdot (p_{1} -p_{2}) + \alpha_{4}(p_{1}-p_2)\cdot
p_{3} )\right )\;.\nonumber \label{p-amp}
\end{eqnarray}

Note that in the $SU(3)$ limit, one has
\begin{eqnarray}
&&T^p(K^{+}(p_{1})K^{+}(p_{2})K^{-}(p_{3})) = T^p(\pi^{+}(p_{1})\pi^{+}(p_{2})\pi^{-}(p_{3}))\;,\nonumber\\
&&T^p(K^{+}(p_{1})\pi^{+}(p_{2})\pi^{-}(p_{3}))=T^p(\pi^{+}(p_{1})K^{+}(p_{2})K^{-}(p_{3}))\;.
\end{eqnarray}
Similarly, one can write down the penguin amplitude $P^p$.

\section{Numerical analysis}

Combining all contributions discussed in previous sections, the total tree $T_t$ and penguin $P_t$ decay amplitudes are then given by
\begin{eqnarray}
T_t = T + T^p + \Delta T \;,\;\;P_t = P + P^p + \Delta P\;. \label{newA}
\end{eqnarray}

The momentum independent contributions have two problems in explaining the data. One problem is that the differences of the branching ratios, that is,
the data do not satisfy the prediction in the approximation given in Eq.(\ref{another1}). The other problem is that except the
ratio of CP asymmetry in $B^+ \to K^+ K^+ K^-$ and $B^+ \to \pi^+ \pi^+ \pi^-$ agree with data, the other ratios predicted  in Eq.(\ref{cpa}) do not agree with data. We now study whether the new total amplitudes in the above can explain the data.

\subsection{Modifications from $T^p$ and $P^p$ only}

If there is no $SU(3)$ breaking contributions, that is  $\Delta T$ and $\Delta P$ vanish, the modifications come from $T^p$ and $P^p$.
In this case, due to contributions from (1) and (2) terms in Eq.(\ref{p-term}), the problem caused by the prediction of Eq.(\ref{another1}) can be fixed. However, note that in this case one still has in the $SU(3)$ limit
\begin{eqnarray}
&&T_t(K^+ K^+ K^-) = T_t (\pi^+ \pi^+ \pi^-)\;,\;\;T_t(\pi^+ K^+ K^-) = T_t(K^+\pi^+ \pi^-) \;,\nonumber\\
&&P_t(K^+ K^+ K^-) = P_t (\pi^+ \pi^+ \pi^-)\;,\;\;P_t(\pi^+ K^+ K^-) = P_t(K^+\pi^+ \pi^-) \;,\nonumber\\
\end{eqnarray}
Because of the above, one has
\begin{eqnarray}
{A_{CP}(\pi^{+}K^{+}K^{-}) \over A_{CP}(K^{+}
 \pi^{+}\pi^{-})}=-{Br(K^{+}
 \pi^{+}\pi^{-}) \over Br(\pi^{+}K^{+}K^{-})}\;,
 \end{eqnarray}
 which is in contradiction with data.

One may wonder if the addition of $T^p$ and $P^p$ can be help to obtain reasonable branching ratios for these decays. We find that this is not the case. Neglecting the masses of $K$, the total decay amplitudes are in the form
\begin{eqnarray}
T_t = a^T + {b^T\over m_B^2}(s+t) + {c^T\over m_B^2}(s-t)\;,\;\;P_t = a^P + {b^P\over m_B^2}(s+t) + {c^P\over m_B^2}(s-t)\;,
\end{eqnarray}
where $s=(p_{2}+p_{3})^2$ and $t=(p_{1}+p_{3})^2$.

The coefficients $a$, $b$ and $c$ can be read off from Eq.s(\ref{leading}) and Eq.(\ref{p-amp}).
We have for all the four decay modes,
\begin{eqnarray}
&&a^{T} = T +{1\over 4} (\alpha^{T}_3 +\alpha^{T}_5)\;,\;\;b^T= {1\over 2}(\alpha^T_{2}-\frac{1}{2}\alpha^T_{3}-\alpha^T_{5}+\frac{1}{2}\alpha^T_{6})\;,
\end{eqnarray}
and $c^T$ is non-zero only for $B^+ \to K^+\pi^+\pi^-,\;\;\pi^+K^+K^-$ decays. It is given by
\begin{eqnarray}
c^{T}={1\over 2}(\alpha^{T}_{1} +\alpha^{T}_{4})\;.
\end{eqnarray}
Similarly, the penguin amplitudes $P^p$ can be written in the same form by replacing $T$ by $P$.

The decay width is then in the form
\begin{eqnarray}
&&\Gamma=\frac{M_B}{512\pi^3}\left(  \mid \tilde a \mid^2 +{2\over 3} (\tilde a \tilde b^{*}+\tilde b\tilde a^{*}) +{1\over 2} \mid \tilde b\mid^2+{1\over 6} \mid \tilde c\mid^2\right )\;, \label{width}
\end{eqnarray}
where
\begin{eqnarray}
&&\tilde a =V_{ub}^{*}V_{ud}a^T+V_{tb}^{*}V_{td}a^P\;,\;\;\tilde b =V_{ub}^{*}V_{ud} b^T+V_{tb}^{*}V_{td} b^P\;,\;\;\tilde c =V_{ub}^{*}V_{ud} c^T+V_{tb}^{*}V_{td} c^P\;.
\end{eqnarray}

From the above formula, we note that the contribution from $\tilde c$  does not interfere with the other two terms. Because this property inclusion of $\tilde c$, if it enhances the branching ratios of $B^+ \to K^+ \pi^+ \pi^-$, it also enhances $B^+ \to \pi^+  K^+ K^-$ compared with
$B^+ \to K^+ K^+ K^-$ and $B^+ \to \pi^+ \pi^+ \pi^-$, respectively. Therefore the addition of contributions from $\tilde c$ does help to improve fit to data which requires enhancement of branching ratio for  $B^+\to K^+ \pi^+ \pi^-$, but reduction for $B^+ \to \pi^+ K^+ K^-$.

We need to include $SU(3)$ breaking $\Delta T$ and $\Delta P$ terms. Before considering both contributions, let us study whether another extreme case, where $T^p$ and $P^p$ are vanishing, but $\Delta T$ and $\Delta P$ are kept non-zero, can explain the data.

\subsection{Modifications from $\Delta T$ and $\Delta P$ only}

For this case, we find it convenient to carry out the analysis by shifting the amplitudes in the following way, indicated by a `` $'$ '' on the amplitudes. For the tree amplitude, redefine $T(\pi^+\pi^+\pi^-)_t = T' = T +\Delta T(\pi^+\pi^+\pi^-)$. Then the other decay amplitudes are
\begin{eqnarray}
&&T(K^+ K^+ K^-)_t = T' +\Delta T'(K^+K^+K^-)\;,\nonumber\\
&&T(\pi^+ K^+ K^-)_t = T' +\Delta T'(\pi^+K^+K^-)\;,\nonumber\\
&&T(K^+ \pi^+ \pi^-)_t = T' +\Delta T'(K^+\pi^+\pi^-)\;,
\end{eqnarray}
where
\begin{eqnarray}
&&\Delta T'(K^+ K^+ K^-) = \Delta T(K^+K^+K^-) - \Delta T(\pi^+\pi^+\pi^-)\;,\nonumber\\
&&\Delta T'(\pi^+ K^+ K^-) = \Delta T(\pi^+K^+K^-) - \Delta T(\pi^+\pi^+\pi^-)\;,\;,\nonumber\\
&&\Delta T'(K^+ \pi^+ \pi^-) = \Delta T(K^+\pi^+\pi^-) - \Delta T(\pi^+\pi^+\pi^-)\;.
\end{eqnarray}

We then have
\begin{eqnarray}
\Delta T'(K^{+}K^{-}K^{+})-\Delta T'(K^{+}\pi^{+}\pi^{-})=\Delta T'(\pi^{+}K^{+}K^{-})\;.
\label{relation2}
\end{eqnarray}
We will take $\Delta T'(K^{+}\pi^{+}\pi^{-})$ and $\Delta T'(\pi^{+}K^{+}K^{-})$ as independent variables in our later ananlysis.
Similarly, one can redefine $P$ and $\Delta P$ to $P'$ and $\Delta P'$.

The complete decay amplitudes, in this case, can be written as
\begin{eqnarray}
&&A(\pi^{+}\pi^{+}\pi^{-})=V_{ub}^{*}V_{ud}T'+V_{tb}^{*}V_{td}P'\;,\\
&&A(K^{+}K^{+}K^{-})=V_{ub}^{*}V_{us}[T'+\Delta
T'(K^{+}K^{+}K^{-})]+V_{tb}^{*}V_{ts}[P'+\Delta
P'(K^{+}K^{+}K^{-})]\;,\nonumber\\
&& A(\pi^{+} K^{+}K^{-})=V_{ub}^{*}V_{ud}[T'+\Delta T'(\pi^{+}
K^{+}K^{-})]+V_{tb}^{*}V_{td}[P'+\Delta
P'(\pi^{+} K^{+}K^{-})]\;,\nonumber\\
&&A(K^{+} \pi^{+}\pi^{-})=V_{ub}^{*}V_{us}[T'+\Delta T'(K^{+}
\pi^{+}\pi^{-})]+V_{tb}^{*}V_{ts}[P'+\Delta P'(K^{+}
\pi^{+}\pi^{-})]\;. \nonumber \label{para}
\end{eqnarray}
The amplitudes for corresponding $B^-$ decays can be obtained by replacing $V_{ub}V_{uq}^*$ and $V_{tb}V_{tq}^*$ in the above by
$V_{ub}^*V_{uq}$ and $V_{tb}^*V_{tq}$, respectively.
One can always choose a convention in which $T'$ is real, and write $P' = P_a + P_b i$, $\Delta T' = \Delta T_a + \Delta T_b i$,  and $\Delta P' = \Delta P_a + \Delta P_b i$.

The KM elements $V_{ij}$ have been very well determined from various experimental data~\cite{PDG}.
In our analysis, we take the KM matrix elements to be known ones and use their central values in the Particle Data Group parameterization~\cite{PDG},
\begin{eqnarray}
&&\theta_{13} =0.0034^{+0.0002}_{-0.0001}\;, \theta_{23}=0.0412^{+0.0011}_{-0.0007}\;, \theta_{12}=0.2273^{+0.0007}_{-0.0007}\; \nonumber\\
&&\delta =1.208^{+0.057}_{-0.039}\;.
\end{eqnarray}

We have seen in section II that the ratio $A_{CP} (\pi^{+} \pi^{+}\pi^{-}) / A_{CP}(K^{+}K^{+}K^{-})$ predicted
by $SU(3)$ conserving momentum independent contributions agree with experimental data well.  If one takes this as an indication that the decay amplitudes for
these two decays obey predictions with just $SU(3)$ momentum independent contributions, and works with the assumption that the $SU(3)$ breaking effects do not modify the relative size of these two decay amplitudes, that is,
$T'(K^+K^+K^-) = T'(\pi^+\pi^+\pi^-)$, one then has
\begin{eqnarray}
&&\Delta T'(K^+K^+K^-) = T'(\pi^+\pi^+\pi^-)=0\;\nonumber\\
&&\Delta T'(K^{+}\pi^{+}\pi^{-})=-\Delta T'(\pi^{+}K^{+}K^{-})\;.
\end{eqnarray} 

In this case, there are  seven parameters for the hadronic matrix elements $T$,
$P_a$, $P_b $, $\Delta T_a$ , $\Delta T_b $,  $\Delta P_a$ and
$\Delta P_b $, to fit the four branching ratios $Br(i)$ and four CP
asymmetries $A_{CP}(i)$. We find that the fit is very good with the
minimal $\chi^2$ to be 0.044. The central values and their 1$\sigma$
allowed ranges are given by,
\begin{eqnarray}
&&T'=(2.70^{+0.14}_{-0.12})\times10^{-5}\;, \nonumber\\
&&P'=P_{a}+P_{b}i=(4.16^{+0.07}_{-0.05})\times 10^{-6}-(7.22^{+1.2}_{-1.0})\times 10^{-7}i\;, \nonumber\\
&&\Delta T'=\Delta T_{a}+\Delta T_{b}i=(-2.04^{+0.21}_{-0.20}) \times 10^{-5}-(2.05^{+0.19}_{-0.17})\times 10^{-5}i\;, 
\nonumber\\
&&\Delta P'=\Delta P_{a}+\Delta P_{b}i=(-1.74^{+0.25}_{-0.24})
\times 10^{-6} -(4.06^{+0.32}_{-0.31}) \times 10^{-6}i\;. \label{in1}
\end{eqnarray}

We list the values for the eight physical observables for the central values for the parameters given above in the first columns of the $A_{CP}$[output] and $Br$[output] in Table II and compare them with data.
From these numbers, one can also see that the fit can be considered as a good one.

\begin{table}[h!]
\begin{centering}
\begin{tabular}{|c||c|c||c|c||}
\hline  $B^+$ decay modes & $A_{CP}$[output]&
$A_{CP}$[data]&Br($10^{-6}$)[output]& Br($10^{-6}$)[data]\\ \hline 
$K^{+}\pi^{+}\pi^{-}$&$0.031,\;\;0.032,\;\; 0.032$& $ 0.032\pm 0.011$& $51.0,\;\;51.0,\;\; 51.1$ &$51.0\pm3.0$\\ \hline 

$K^{+}K^{+}K^{-}$&$-0.042,\;\;-0.043,\;\;-0.043$ &$-0.043\pm0.012$ &$33.9,\;\;34.1,\;\;33.9$ &$34.0\pm1.0$\\ \hline 

$\pi^{+}\pi^{+}\pi^{-}$ & $0.120,\;\;0.117,\;\;0.118$&$0.117 \pm 0.024$ &$15.2,\;\;15.2,\;\;15.2$&$15.2 \pm1.4$ \\ \hline 

$\pi^+ K^{+}K^{-}$&$-0.142,\;\;-0.143,\;\;-0.140$& $-0.141 \pm 0.044 $ &$5.0,\;\;5.0,\;\; 5.0$& $5.0 \pm 0.7$\\ \hline
\end{tabular} \label{results}
\caption{Comparison of experimental data and fit values with $SU(3)$ breaking effects. The first, second and third columns for outputs are for the cases of input parameters from Eq.(\ref{in1}), Eq.(\ref{in2}) and Eq.(\ref{in3}), respectively. }
\end{centering}\end{table}

The ratio for $A_{CP}(\pi^{+}\pi^{+}\pi^{-}) /
A_{CP}(K^{+}K^{+}K^{-})$ is predicted to be $-2.2\pm 0.2$. 
This agrees with $-2.7\pm 0.9$ determined from data within error bars.  The central values are, however, different. 
If this persists with more data, one may need to keep $\Delta T'(K^+K^+K^-)$ to be non-zero to fit data. 
For example with the following value for the amplitudes
\begin{eqnarray}
&&T'=1.7\times10^{-5}\;,\;\;P'=-4.5\times10^{-6}-5.7\times10^{-7}i\;,\nonumber\\
&&\Delta T'(K^{+}\pi^{+}\pi^{-})=3.3\times10^{-5}+4.2\times10^{-5}i\;,\nonumber\\
&&\Delta P'(K^{+}\pi^{+}\pi^{-})=-3.3\times10^{-6}-4.4\times10^{-6}i\;,\nonumber\\
&&\Delta T'(\pi^{+}K^{+}K^{-})=-3.4\times10^{-5}-1.8\times10^{-5}i\;,\nonumber\\
&&\Delta P'(\pi^{+}
K^{+}K^{-})=-4.5\times10^{-6}-5.7\times10^{-7}i\;,\nonumber \label{in2}
\end{eqnarray}
we can have $A_{CP}(\pi^{+}\pi^{+}\pi^{-}) /
A_{CP}(K^{+}K^{+}K^{-})=-2.7$ coincident with the central value of the data.  The corresponding 
outputs for other observables are give in the second columns of Table II for outputs.

The above analysis shows that it is possible to have consistent solution with the branching ratios and CP asymmetries provided that the $SU(3)$ breaking effects are large. 

\subsection{Modifications from both $SU(3)$ breaking and derivative terms}

We now carry out an analysis by including both $\Delta T$ ($\Delta P$), and $T^p(P^p)$ terms to see how things will change. In particular one wonders if including $T^p (P^p)$ contributions, one can fit data well with small $SU(3)$ breaking effects. 

If finite $m_K$ is kept, the derivative terms also contain $SU(3)$ breaking terms. These additional terms satisfy Eq.({\ref{relation2}) and by appropriate redefinition of the amplitudes, one can group them into $T (P)$, $\Delta T' (\Delta P')$ terms. The total amplitudes can be written as
\begin{eqnarray}
&&T_t = a^T  + {b^T\over m_B^2}(s+t) + {c^T\over m_B^2}(s-t) + \Delta a^T\;,\nonumber\\
&&P_t = a^P + {b^P\over m_B^2}(s+t) + {c^P\over m_B^2}(s-t) + \Delta a^P\;,
\end{eqnarray}
where
\begin{eqnarray}
&&\Delta a^T(\pi^+\pi^+\pi^-) = 0\;,\nonumber\\
&&\Delta a^T(K^+K^+ K^-) = \Delta T'(K^+K^+K^-)\;,\nonumber\\
&&\Delta a^T(K^+\pi^+ \pi^-) = \Delta T'(K^+\pi^+\pi^-)\;,\nonumber\\
&&\Delta a^T(\pi^+K^+K^-) = \Delta T'(\pi^+ K^+ K^-)\;.
\end{eqnarray}

Eq.(\ref{width}) can still be used for decay width calculation by replacing $a^T $ and $a^P$ by
$a^T +\Delta a^T$ and $a^P+\Delta a^P$. We find that the situation does not change much compared with including
just $\Delta T'$ and $\Delta P'$ corrections. We cannot find solutions with $SU(3)$ breaking terms to be much smaller than the $SU(3)$ conserving terms. The reasons are that the momentum dependent contributions contribute to all 
$T' (P')$, $\Delta T' (\Delta P')$ and also have  pieces $\tilde c^T (\tilde c^P)$. The former two contributions can be absorbed into $T' (P')$ and $\Delta T' (\Delta (P')$, and therefore the net results cannot be improved by these two terms. The pieces 
$\tilde c^T (\tilde c^P)$ do not interfere with the other two terms. Because this property, as already pointed out before,  inclusion of $\tilde c$ only enhances the branching ratios of $B^+ \to K^+ \pi^+ \pi^-$ and $B^+ \to \pi^+  K^+ K^-$ compared with $B^+ \to K^+ K^+ K^-$ and $B^+ \to \pi^+ \pi^+ \pi^-$, respectively. The addition of contributions from $\tilde c$ does not help to improve fit for small $\Delta T' (\Delta P')$. 

We have made a scan of parameter space and have not found solutions with the magnitude of $\Delta T' (\Delta P')$ to be a few times smaller than the magnitudes of $T'(P')$. But the magnitudes of $T^p$ and $P^p$ are not necessarily to be small (in the limit they are zero, the situation goes back to the case discussed int he previous sub-section). For example, with
\begin{eqnarray}
&&a^{T}=1.3\times10^{-5}\;,\;\;a^{P}=1.1\times10^{-6}+5.7\times10^{-6}i\;,\nonumber\\
&&b^{T}=1.1\times10^{-5}-1.2\times10^{-5}i\;,\;\;b^{P}=7.0\times10^{-6}-2.1\times10^{-5}i\;,\nonumber\\
&&c^{T}=2.4\times10^{-5}+4.3\times10^{-5}i\;,\;\;c^{P}=1.5\times10^{-5}+3.4\times10^{-5}i\;,\nonumber\\
&&\Delta a^{T}(K^{+}\pi^{+}\pi^{-})=-1.2\times10^{-5}-1.1\times10^{-5}i\;,\\
&&\Delta a^{P}(K^{+}\pi^{+}\pi^{-})=3.0\times10^{-6}-1.4\times10^{-6}i\;,\nonumber\\
&&\Delta a^{T}(K^{+}K^{-}\pi^{+})=-1.4\times10^{-5}+1.4\times10^{-5}i\;,\nonumber\\
&&\Delta a^{P}(K^{+}K^{-}\pi^{+})=-3.8\times10^{-6}+5.0\times10^{-6}i\;,\nonumber \label{in3}
\end{eqnarray}
we can obtain results in good agreement with the experiment data as in the third columns for outputs in Table II.

\section{Summary}

We have studied CP violation in charged 3-body $B^+$ decays,
$B^+\to \pi^+ K^+ K^-$, $B^+\to \pi^+\pi^+\pi^-$, $B^+\to K^+ K^+ K^- $ and $B^+\to K^+\pi^+\pi^-$,
using flavor $SU(3)$ symmetry. Several contributions are considered including $SU(3)$ conserving momentum independent amplitudes ($T'$ and $P'$), and momentum dependent amplitudes ($T^p$ and $P^p$), and also  $SU(3)$ breaking amplitudes ($\Delta T'$ and $\Delta P'$). We have studied how to constructed non-resonant contributions to these amplitudes. 

With only $SU(3)$ conserving momentum independent contributions,  one would have branching ratios and CP violating rate differences satisfy, $Br(\pi^+\pi^+\pi^-) = 2Br(\pi^+ K^+ K^-)$,
$Br(K^+K^+K^-) = 2 Br(K^+\pi^+\pi^-)$, and $\Delta_{CP}(\pi^+\pi^+\pi^-) = 2\Delta_{CP}(\pi^+ K^+K^-) = -
\Delta_{CP}(K^+K^+K^-) = -2\Delta_{CP}(K^+\pi^+\pi^-)$. The LHCb data do not exhibit the expected pattern for the branching ratios. The rate differences for $B^+\to \pi^+\pi^+\pi^-$ and $B^+\to K^+ K^+ K^-$ satisfy the relation between $\Delta S =0$ and $\Delta S=1$ well, but the other two do not, with the CP asymmetries having different signs than expected.  One needs to go beyond the $SU(3)$ limit description for these decays.

With $SU(3)$ conserving momentum dependent amplitudes added, the degeneracy between the amplitudes for $A(\pi^+\pi^+\pi^-)$ and $A(\pi^+ K^+ K^-)$, and $A(K^+K^+K^-)$ and $A(K^+\pi^+\pi^-)$ can be lifted by a new piece of contribution (the $\tilde c^{T,P}$ contributions). Because this new contribution does not interfere with the other contributions if it enhances the branching ratios of $B^+ \to K^+ \pi^+ \pi^-$, it also enhances $B^+ \to \pi^+  K^+ K^-$ compared with
$B^+ \to K^+ K^+ K^-$ and $B^+ \to \pi^+ \pi^+ \pi^-$, respectively. This does not help to improve fit to data which requires enhancement of branching ratio for  $B^+\to K^+ \pi^+ \pi^-$, but reduction for $B^+ \to \pi^+ K^+ K^-$. 

With $SU(3)$ breaking contributions from quark mass differences included, we find that the experimental data can be explained  provided that the breaking effects are large, comparable strength with $SU(3)$ conserving amplitudes. One wonders if such a large modification is reasonable. One might expect that the
breaking effects should be of order a few decades percent. However, one notices that contributions to each of the $\Delta T'$ and $\Delta P'$ have several terms. Even though each of the terms
is only at most 10\% , if they contribute constructively, the total effects can be much larger. This might be what is happening for the decays we are considering here. At phenomenological level, this
cannot be fully understood. Model calculations do have found large corrections~\cite{hycheng}. One hopes that this problem can be understood better from lattice calculations in the future.

\begin{acknowledgments}

XG would like to thank Hai-Yang Cheng and Jusak Tandean for useful comments.
The work was supported in part by MOE Academic Excellent Program (Grant No: 102R891505) and NSC of ROC, and in part by NNSF(Grant No:11175115) and Shanghai Science and Technology Commission (Grant No: 11DZ2260700) of PRC.

\end{acknowledgments}

\appendix

\section{$SU(3)$ breaking terms}

Following the description of constructing the $SU(3)$ breaking corrections in the text, the $SU(3)$ breaking tree amplitude $T(q)$ can be expressed as
\begin{eqnarray}
\Delta T(q)&&=a^T_{1}(\overline{3})B_{i}H^{a}(\overline{3})W^{i}_{a}M^{j}_{k}M^{k}_{l}M^{l}_{j}+a^T_{2}(\overline{3})B_{i}H^{i}(\overline{3})M^{j}_{k}M^{k}_{l}M^{a}_{j}W^{l}_{a}\nonumber\\
&&+b^T_{1}(\overline{3})H^{i}(\overline{3})W^{a}_{i}M^{j}_{a}B_{j}M^{k}_{l}M^{l}_{k}+b^T_{2}(\overline{3})H^{i}(\overline{3})M^{j}_{i}B_{a}W^{a}_{j}M^{k}_{l}M^{l}_{k}\nonumber\\
&&+b^T_{3}(\overline{3})H^{i}(\overline{3})M^{j}_{i}B_{j}M^{k}_{l}M^{a}_{k}W^{l}_{a}\nonumber\\
&&+c^T_{1}(\overline{3})H^{a}(\overline{3})W^{i}_{a}M^{l}_{i}M^{j}_{l}M^{k}_{j}B_{k}+c^T_{2}(\overline{3})H^{i}(\overline{3})M^{l}_{i}M^{a}_{l}W^{j}_{a}M^{k}_{j}B_{k}\nonumber\\
&&+c^T_{3}(\overline{3})H^{i}(\overline{3})M^{l}_{i}M^{j}_{l}M^{a}_{j}W^{k}_{a}B_{k}+c^T_{4}(\overline{3})H^{i}(\overline{3})M^{l}_{i}W^{a}_{l}M^{j}_{a}M^{k}_{j}B_{k}\nonumber\\
&&+a^T_{1}(6)B_{a}W^{a}_{i}H^{ij}_{k}(6)M^{k}_{j}M^{l}_{n}M^{n}_{l}+a^T_{2}(6)B_{i}H^{ia}_{k}(6)W^{j}_{a}M^{k}_{j}M^{l}_{n}M^{n}_{l}\nonumber\\
&&+a^T_{3}(6)B_{i}H^{ij}_{k}(6)W^{k}_{a}M^{a}_{j}M^{l}_{n}M^{n}_{l}+a^T_{4}(6)B_{i}H^{ij}_{k}(6)M^{k}_{j}M^{l}_{n}M^{a}_{l}W^{n}_{a}\nonumber\\
&&+b^T_{1}(6)B_{a}W^{a}_{i}H^{ij}_{k}(6)M^{k}_{l}M^{l}_{n}M^{n}_{j}+b^T_{2}(6)B_{i}H^{ia}_{k}(6)W^{j}_{a}M^{k}_{l}M^{l}_{n}M^{n}_{j}\nonumber\\
&&+b^T_{3}(6)B_{i}H^{ij}_{k}(6)W^{k}_{a}M^{a}_{l}M^{l}_{n}M^{n}_{j}+b^T_{4}(6)B_{i}H^{ij}_{k}(6)W^{a}_{l}M^{k}_{a}M^{l}_{n}M^{n}_{j}\nonumber\\
&&+b^T_{5}(6)B_{i}H^{ij}_{k}(6)W^{a}_{n}M^{k}_{l}M^{l}_{a}M^{n}_{j}\nonumber\\
&&+c^T_{1}(6)B_{i}H^{jk}_{l}(6)W^{i}_{a}M^{a}_{j}M^{n}_{k}M^{l}_{n}+c^T_{2}(6)B_{i}H^{jk}_{l}(6)W^{a}_{j}M^{i}_{a}M^{n}_{k}M^{l}_{n}\nonumber\\
&&+c^T_{3}(6)B_{i}H^{jk}_{l}(6)M^{i}_{j}W^{a}_{k}M^{n}_{a}M^{l}_{n}+c^T_{4}(6)B_{i}H^{jk}_{l}(6)M^{i}_{j}M^{n}_{k}W^{l}_{a}M^{a}_{n}\nonumber\\
&&+c^T_{5}(6)B_{i}H^{jk}_{l}(6)M^{i}_{j}M^{n}_{k}M^{l}_{a}W^{a}_{n}\nonumber\\
&&+d^T_{1}(6)B_{a}W^{a}_{i}H^{jk}_{l}(6)M^{i}_{n}M^{l}_{j}M^{n}_{k}+d^T_{2}(6)B_{i}H^{ak}_{l}(6)W^{j}_{a}M^{i}_{n}M^{l}_{j}M^{n}_{k}\nonumber\\
&&+d^T_{3}(6)B_{i}H^{ja}_{l}(6)W^{k}_{a}M^{i}_{n}M^{l}_{j}M^{n}_{k}+d^T_{4}(6)B_{i}H^{jk}_{a}W^{a}_{l}(6)M^{i}_{n}M^{l}_{j}M^{n}_{k}\nonumber\\
&&+d^T_{5}(6)B_{i}H^{jk}_{l}(6)W^{a}_{n}M^{i}_{a}M^{l}_{j}M^{n}_{k}\nonumber\\
&&+a^T_{1}(\overline {15})B_{a}W^{a}_{i}H^{ij}_{k}(\overline {15})M^{k}_{j}M^{l}_{n}M^{n}_{l}+a^T_{2}(\overline {15})B_{i}H^{ia}_{k}(\overline {15})W^{j}_{a}M^{k}_{j}M^{l}_{n}M^{n}_{l}\nonumber\\
&&+a^T_{3}(\overline {15})B_{i}H^{ij}_{k}(\overline {15})W^{k}_{a}M^{a}_{j}M^{l}_{n}M^{n}_{l}+a^T_{4}(\overline {15})B_{i}H^{ij}_{k}(\overline {15})M^{k}_{j}M^{l}_{n}M^{a}_{l}W^{n}_{a}\nonumber\\
&&+b^T_{1}(\overline {15})B_{a}W^{a}_{i}H^{ij}_{k}(\overline {15})M^{k}_{l}M^{l}_{n}M^{n}_{j}+b^T_{2}(\overline {15})B_{i}H^{ia}_{k}(\overline {15})W^{j}_{a}M^{k}_{l}M^{l}_{n}M^{n}_{j}\nonumber\\
&&+b^T_{3}(\overline {15})B_{i}H^{ij}_{k}(\overline {15})W^{k}_{a}M^{a}_{l}M^{l}_{n}M^{n}_{j}+b^T_{4}(\overline {15})B_{i}H^{ij}_{k}(\overline {15})W^{a}_{l}M^{k}_{a}M^{l}_{n}M^{n}_{j}\nonumber\\
&&+b^T_{5}(\overline {15})B_{i}H^{ij}_{k}(\overline {15})W^{a}_{n}M^{k}_{l}M^{l}_{a}M^{n}_{j}\nonumber\\
&&+c^T_{1}(\overline {15})B_{i}H^{jk}_{l}(\overline {15})W^{i}_{a}M^{a}_{j}M^{n}_{k}M^{l}_{n}+c^T_{2}(\overline {15})B_{i}H^{jk}_{l}(\overline {15})W^{a}_{j}M^{i}_{a}M^{n}_{k}M^{l}_{n}\nonumber\\
&&+c^T_{3}(\overline {15})B_{i}H^{jk}_{l}(\overline {15})M^{i}_{j}W^{a}_{k}M^{n}_{a}M^{l}_{n}+c^T_{4}(\overline {15})B_{i}H^{jk}_{l}(\overline {15})M^{i}_{j}M^{n}_{k}W^{l}_{a}M^{a}_{n}\nonumber\\
&&+c^T_{5}(\overline {15})B_{i}H^{jk}_{l}(\overline {15})M^{i}_{j}M^{n}_{k}M^{l}_{a}W^{a}_{n}\nonumber\\
&&+d^T_{1}(\overline {15})B_{a}W^{a}_{i}H^{jk}_{l}(\overline {15})M^{i}_{n}M^{l}_{j}M^{n}_{k}+d^T_{2}(\overline {15})B_{i}H^{ak}_{l}(\overline {15})W^{j}_{a}M^{i}_{n}M^{l}_{j}M^{n}_{k}\nonumber\\
&&+d^T_{3}(\overline {15})B_{i}H^{ja}_{l}(\overline {15})W^{k}_{a}M^{i}_{n}M^{l}_{j}M^{n}_{k}+d^T_{4}(\overline {15})B_{i}H^{jk}_{a}W^{a}_{l}(\overline {15})M^{i}_{n}M^{l}_{j}M^{n}_{k}\nonumber\\
&&+d^T_{5}(\overline {15})B_{i}H^{jk}_{l}(\overline {15})W^{a}_{n}M^{i}_{a}M^{l}_{j}M^{n}_{k}\nonumber\\
\end{eqnarray}
The penguin amplitude $P(q)$ can be obtained by replacing $T$ by $P$.

\section{ Terms with two derivatives}

In this appendix, we list independent $SU(3)$ invariant amplitude $T^p$ with two derivatives in the following.
\begin{eqnarray}
T^p(q) &=&a^{\prime}(\overline{3})_{1}(\partial_{\mu}B_{i})H^{i}(\overline{3})(\partial^{\mu}M^{j}_{k})M^{k}_{l}M^{l}_{j}+ a^{\prime}(\overline{3})_{2}(\partial_{\mu}B_{i})H^{i}(\overline{3})M^{j}_{k}(\partial^{\mu}M^{k}_{l})M^{l}_{j} \nonumber\\
&+&a^{\prime}(\overline{3})_{3}(\partial_{\mu}B_{i})H^{i}(\overline{3})M^{j}_{k}M^{k}_{l}(\partial^{\mu}M^{l}_{j})+ a^{\prime\prime}(\overline{3})_{1}B_{i}H^{i}(\overline{3})(\partial^{\mu}M^{j}_{k})(\partial_{\mu}M^{k}_{l})M^{l}_{j} \nonumber\\
&+&a^{\prime\prime}(\overline{3})_{2}B_{i}H^{i}(\overline{3})(\partial^{\mu}M^{j}_{k})M^{k}_{l}(\partial_{\mu}M^{l}_{j})+ a^{\prime\prime}(\overline{3})_{3}B_{i}H^{i}(\overline{3})M^{j}_{k}(\partial^{\mu}M^{k}_{l})(\partial_{\mu}M^{l}_{j})  \nonumber\\
&+&b^{\prime}(\overline{3})_{1}(\partial_{\mu}B_{j})H^{i}(\overline{3})(\partial^{\mu}M^{j}_{i})M^{k}_{l}M^{l}_{k}+  b^{\prime}(\overline{3})_{2}(\partial_{\mu}B_{j})H^{i}(\overline{3})M^{j}_{i}(\partial^{\mu}M^{k}_{l})M^{l}_{k} \nonumber\\
&+&b^{\prime}(\overline{3})_{3}(\partial_{\mu}B_{j})H^{i}(\overline{3})M^{j}_{i}M^{k}_{l}(\partial^{\mu}M^{l}_{k})+ b^{\prime\prime}(\overline{3})_{1}B_{j}H^{i}(\overline{3})(\partial^{\mu}M^{j}_{i})(\partial_{\mu}M^{k}_{l})M^{l}_{k} \nonumber\\
&+&b^{\prime\prime}(\overline{3})_{2}B_{j}H^{i}(\overline{3})(\partial^{\mu}M^{j}_{i})M^{k}_{l}(\partial_{\mu}M^{l}_{k})+ b^{\prime\prime}(\overline{3})_{3}B_{j}H^{i}(\overline{3})M^{j}_{i}(\partial^{\mu}M^{k}_{l})(\partial_{\mu}M^{l}_{k})  \nonumber\\
&+&c^{\prime}(\overline{3})_{1}(\partial_{\mu}B_{k})H^{i}(\overline{3})(\partial^{\mu}M^{l}_{i})M^{j}_{l}M^{k}_{j}+ c^{\prime}(\overline{3})_{2}(\partial_{\mu}B_{k})H^{i}(\overline{3})M^{l}_{i}(\partial^{\mu}M^{j}_{l})M^{k}_{j} \nonumber\\
&+&c^{\prime}(\overline{3})_{3}(\partial_{\mu}B_{k})H^{i}(\overline{3})M^{l}_{i}M^{j}_{l}(\partial^{\mu}M^{k}_{j})+ c^{\prime\prime}(\overline{3})_{1}B_{k}H^{i}(\overline{3})(\partial^{\mu}M^{l}_{i})(\partial_{\mu}M^{j}_{l})M^{k}_{j}  \nonumber\\
&+&c^{\prime\prime}(\overline{3})_{2}B_{k}H^{i}(\overline{3})(\partial^{\mu}M^{l}_{i})M^{j}_{l}(\partial_{\mu}M^{k}_{j})+ c^{\prime\prime}(\overline{3})_{3}B_{k}H^{i}(\overline{3})M^{l}_{i}(\partial^{\mu}M^{j}_{l})(\partial_{\mu}M^{k}_{j})   \nonumber\\
&+&a^{\prime}(6)_{1}(\partial_{\mu}B_{i})H^{ij}_{k}(6)(\partial^{\mu}M^{k}_{j})M^{l}_{n}M^{n}_{l}+ a^{\prime}(6)_{2}(\partial_{\mu}B_{i})H^{ij}_{k}(6)M^{k}_{j}(\partial^{\mu}M^{l}_{n})M^{n}_{l} \nonumber\\
&+&a^{\prime}(6)_{3}(\partial_{\mu}B_{i})H^{ij}_{k}(6)M^{k}_{j}M^{l}_{n}(\partial^{\mu}M^{n}_{l})+ a^{\prime\prime}(6)_{1}B_{i}H^{ij}_{k}(6)(\partial^{\mu}M^{k}_{j})(\partial_{\mu}M^{l}_{n})M^{n}_{l} \nonumber\\
&+&a^{\prime\prime}(6)_{2}B_{i}H^{ij}_{k}(6)(\partial^{\mu}M^{k}_{j})M^{l}_{n}(\partial_{\mu}M^{n}_{l})+ a^{\prime\prime}(6)_{3}B_{i}H^{ij}_{k}(6)M^{k}_{j}(\partial^{\mu}M^{l}_{n})(\partial_{\mu}M^{n}_{l}) \nonumber\\
&+&b^{\prime}(6)_{1}(\partial_{\mu}B_{i})H^{ij}_{k}(6)(\partial^{\mu}M^{k}_{l})M^{l}_{n}M^{n}_{j}+ b^{\prime}(6)_{2}(\partial_{\mu}B_{i})H^{ij}_{k}(6)M^{k}_{l}(\partial^{\mu}M^{l}_{n})M^{n}_{j} \nonumber\\
&+&b^{\prime}(6)_{3}(\partial_{\mu}B_{i})H^{ij}_{k}(6)M^{k}_{l}M^{l}_{n}(\partial^{\mu}M^{n}_{j})+ b^{\prime\prime}(6)_{1}B_{i}H^{ij}_{k}(6)(\partial^{\mu}M^{k}_{l})(\partial_{\mu}M^{l}_{n})M^{n}_{j} \nonumber\\
&+&b^{\prime\prime}(6)_{2}B_{i}H^{ij}_{k}(6)(\partial^{\mu}M^{k}_{l})M^{l}_{n}(\partial_{\mu}M^{n}_{j})+ b^{\prime\prime}(6)_{3}B_{i}H^{ij}_{k}(6)M^{k}_{l}(\partial^{\mu}M^{l}_{n})(\partial_{\mu}M^{n}_{j}) \nonumber\\
&+&c^{\prime}(6)_{1}(\partial_{\mu}B_{i})H^{ik}_{l}(6)(\partial^{\mu}M^{i}_{j})M^{n}_{k}M^{l}_{n}+ c^{\prime}(6)_{2}(\partial_{\mu}B_{i})H^{jk}_{l}(6)M^{i}_{j}(\partial^{\mu}M^{n}_{k})M^{l}_{n} \nonumber\\
&+&c^{\prime}(6)_{3}(\partial_{\mu}B_{i})H^{jk}_{l}(6)M^{i}_{j}M^{n}_{k}(\partial^{\mu}M^{l}_{n})+ c^{\prime\prime}(6)_{1}B_{i}H^{jk}_{l}(6)(\partial^{\mu}M^{i}_{j})(\partial_{\mu}M^{n}_{k})M^{l}_{n} \nonumber\\
&+&c^{\prime\prime}(6)_{2}B_{i}H^{jk}_{l}(6)(\partial^{\mu}M^{i}_{j})M^{n}_{k}(\partial_{\mu}M^{l}_{n})+ c^{\prime\prime}(6)_{3}B_{i}H^{jk}_{l}(6)M^{i}_{j}(\partial^{\mu}M^{n}_{k})(\partial_{\mu}M^{l}_{n}) \nonumber\\
&+&d^{\prime}(6)_{1}(\partial_{\mu}B_{i})H^{ik}_{l}(6)(\partial^{\mu}M^{i}_{n})M^{l}_{j}M^{n}_{k}+ d^{\prime}(6)_{2}(\partial_{\mu}B_{i})H^{jk}_{l}(6)M^{i}_{n}(\partial^{\mu}M^{l}_{j})M^{n}_{k} \nonumber\\
&+&d^{\prime}(6)_{3}(\partial_{\mu}B_{i})H^{jk}_{l}(6)M^{i}_{n}M^{l}_{j}(\partial^{\mu}M^{n}_{k})+ d^{\prime\prime}(6)_{1}B_{i}H^{jk}_{l}(6)(\partial^{\mu}M^{i}_{n})(\partial_{\mu}M^{l}_{j})M^{n}_{k} \nonumber\\
&+&d^{\prime\prime}(6)_{2}B_{i}H^{jk}_{l}(6)(\partial^{\mu}M^{i}_{n})M^{l}_{j}(\partial_{\mu}M^{n}_{k})+ d^{\prime\prime}(6)_{3}B_{i}H^{jk}_{l}(6)M^{i}_{n}(\partial^{\mu}M^{l}_{j})(\partial_{\mu}M^{n}_{k}) \\
&+&a^{\prime}(\overline{15})_{1}(\partial_{\mu}B_{i})H^{ij}_{k}(\overline{15})(\partial^{\mu}M^{k}_{j})M^{l}_{n}M^{n}_{l}+ a^{\prime}(\overline{15})_{2}(\partial_{\mu}B_{i})H^{ij}_{k}(\overline{15})M^{k}_{j}(\partial^{\mu}M^{l}_{n})M^{n}_{l} \nonumber\\
&+&a^{\prime}(\overline{15})_{3}(\partial_{\mu}B_{i})H^{ij}_{k}(\overline{15})M^{k}_{j}M^{l}_{n}(\partial^{\mu}M^{n}_{l})+ a^{\prime\prime}(\overline{15})_{1}B_{i}H^{ij}_{k}(\overline{15})(\partial^{\mu}M^{k}_{j})(\partial_{\mu}M^{l}_{n})M^{n}_{l} \nonumber\\
&+&a^{\prime\prime}(\overline{15})_{2}B_{i}H^{ij}_{k}(\overline{15})(\partial^{\mu}M^{k}_{j})M^{l}_{n}(\partial_{\mu}M^{n}_{l})+ a^{\prime\prime}(\overline{15})_{3}B_{i}H^{ij}_{k}(\overline{15})M^{k}_{j}(\partial^{\mu}M^{l}_{n})(\partial_{\mu}M^{n}_{l}) \nonumber\\
&+&b^{\prime}(\overline{15})_{1}(\partial_{\mu}B_{i})H^{ij}_{k}(\overline{15})(\partial^{\mu}M^{k}_{l})M^{l}_{n}M^{n}_{j}+ b^{\prime}(\overline{15})_{2}(\partial_{\mu}B_{i})H^{ij}_{k}(\overline{15})M^{k}_{l}(\partial^{\mu}M^{l}_{n})M^{n}_{j} \nonumber\\
&+&b^{\prime}(\overline{15})_{3}(\partial_{\mu}B_{i})H^{ij}_{k}(\overline{15})M^{k}_{l}M^{l}_{n}(\partial^{\mu}M^{n}_{j})+ b^{\prime\prime}(\overline{15})_{1}B_{i}H^{ij}_{k}(\overline{15})(\partial^{\mu}M^{k}_{l})(\partial_{\mu}M^{l}_{n})M^{n}_{j} \nonumber\\
&+&b^{\prime\prime}(\overline{15})_{2}B_{i}H^{ij}_{k}(\overline{15})(\partial^{\mu}M^{k}_{l})M^{l}_{n}(\partial_{\mu}M^{n}_{j})+ b^{\prime\prime}(\overline{15})_{3}B_{i}H^{ij}_{k}(\overline{15})M^{k}_{l}(\partial^{\mu}M^{l}_{n})(\partial_{\mu}M^{n}_{j}) \nonumber\\
&+&c^{\prime}(\overline{15})_{1}(\partial_{\mu}B_{i})H^{ik}_{l}(\overline{15})(\partial^{\mu}M^{i}_{j})M^{n}_{k}M^{l}_{n}+ c^{\prime}(\overline{15})_{2}(\partial_{\mu}B_{i})H^{jk}_{l}(\overline{15})M^{i}_{j}(\partial^{\mu}M^{n}_{k})M^{l}_{n} \nonumber\\
&+&c^{\prime}(\overline{15})_{3}(\partial_{\mu}B_{i})H^{jk}_{l}(\overline{15})M^{i}_{j}M^{n}_{k}(\partial^{\mu}M^{l}_{n})+ c^{\prime\prime}(\overline{15})_{1}B_{i}H^{jk}_{l}(\overline{15})(\partial^{\mu}M^{i}_{j})(\partial_{\mu}M^{n}_{k})M^{l}_{n} \nonumber\\
&+&c^{\prime\prime}(\overline{15})_{2}B_{i}H^{jk}_{l}(\overline{15})(\partial^{\mu}M^{i}_{j})M^{n}_{k}(\partial_{\mu}M^{l}_{n})+ c^{\prime\prime}(\overline{15})_{3}B_{i}H^{jk}_{l}(\overline{15})M^{i}_{j}(\partial^{\mu}M^{n}_{k})(\partial_{\mu}M^{l}_{n}) \nonumber\\
&+&d^{\prime}(\overline{15})_{1}(\partial_{\mu}B_{i})H^{ik}_{l}(\overline{15})(\partial^{\mu}M^{i}_{n})M^{l}_{j}M^{n}_{k}+ d^{\prime}(\overline{15})_{2}(\partial_{\mu}B_{i})H^{jk}_{l}(\overline{15})M^{i}_{n}(\partial^{\mu}M^{l}_{j})M^{n}_{k} \nonumber\\
&+&d^{\prime}(\overline{15})_{3}(\partial_{\mu}B_{i})H^{jk}_{l}(\overline{15})M^{i}_{n}M^{l}_{j}(\partial^{\mu}M^{n}_{k})+ d^{\prime\prime}(\overline{15})_{1}B_{i}H^{jk}_{l}(\overline{15})(\partial^{\mu}M^{i}_{n})(\partial_{\mu}M^{l}_{j})M^{n}_{k} \nonumber\\
&+&d^{\prime\prime}(\overline{15})_{2}B_{i}H^{jk}_{l}(\overline{15})(\partial^{\mu}M^{i}_{n})M^{l}_{j}(\partial_{\mu}M^{n}_{k})+ d^{\prime\prime}(\overline{15})_{3}B_{i}H^{jk}_{l}(\overline{15})M^{i}_{n}(\partial^{\mu}M^{l}_{j})(\partial_{\mu}M^{n}_{k})\;. \nonumber
\end{eqnarray}

Expanding the above, we obtain the expressions for $\alpha_i$ in the following.

\begin{eqnarray}
&& {\alpha_{1}\over m_B^2}~=~-c^{\prime}(6)_{2}+c^{\prime}(\overline{15})_{1} \nonumber\\
&&{\alpha_{2}\over m_B^2}~=~b^{\prime}(\overline{3})_{2}+b^{\prime}(\overline{3})_{3}+c^{\prime}(\overline{3})_{2}+a^{\prime}(6)_{2}+a^{\prime}(6)_{3}+b^{\prime}(6)_{2}-c^{\prime}(6)_{2}-d^{\prime}(6)_{3} \nonumber\\
&&\qquad\qquad +3a^{\prime}(\overline{15})_{2}+3a^{\prime}(\overline{15})_{3}+3b^{\prime}(\overline{15})_{2}+3c^{\prime}(\overline{15})_{2}+3d^{\prime}(\overline{15})_{3}-2c^{\prime}(\overline{15})_{3}  \nonumber\\
&&{\alpha_{3}\over m_B^2}~=~{1 \over 2} \{ [2b^{\prime}(\overline{3})_{1}+c^{\prime}(\overline{3})_{1}+2a^{\prime}(6)_{1}+b^{\prime}(6)_{3}-d^{\prime}(6)_{2} \nonumber\\
&&\qquad\qquad
+6a^{\prime}(\overline{15})_{1}+3b^{\prime}(\overline{15})_{3}+c^{\prime}(\overline{15})_{1}-c^{\prime}(\overline{15})_{2}+3d^{\prime}(\overline{15})_{2}
]\nonumber\\
&&\qquad\qquad +[b^{\prime}(\overline{3})_{2}+b^{\prime}(\overline{3})_{3}+c^{\prime}(\overline{3})_{3}+a^{\prime}(6)_{2}+a^{\prime}(6)_{3}+b^{\prime}(6)_{1}-c^{\prime}(6)_{1}-c^{\prime}(6)_{3} \nonumber\\
&&\qquad\qquad
-d^{\prime}(6)_{1}+3a^{\prime}(\overline{15})_{2}+3a^{\prime}(\overline{15})_{3}+3b^{\prime}(\overline{15})_{1}-c^{\prime}(\overline{15})_{2}+3c^{\prime}(\overline{15})_{3}
+3d^{\prime}(\overline{15})_{1}] \}  \nonumber\\
&&{\alpha_{4}\over m_B^2}~=~c^{\prime\prime}(6)_{2}-c^{\prime\prime}(\overline{15})_{3}   \nonumber\\
&&{\alpha_{5}\over m_B^2}~=~b^{\prime\prime}(\overline{3})_{1}+b^{\prime\prime}(\overline{3})_{2}+c^{\prime\prime}(\overline{3})_{2}+a^{\prime\prime}(6)_{1}+a^{\prime\prime}(6)_{2}+b^{\prime\prime}(6)_{2}-c^{\prime\prime}(6)_{2} \nonumber\\
&&\qquad\qquad -d^{\prime\prime}(6)_{1}+3a^{\prime\prime}(\overline{15})_{1}+3a^{\prime\prime}(\overline{15})_{2}+3b^{\prime\prime}(\overline{15})_{2}-2c^{\prime\prime}(\overline{15})_{1}+3c^{\prime\prime}(\overline{15})_{2}+3d^{\prime\prime}(\overline{15})_{1} \nonumber\\
&&{\alpha_{6}\over m_B^2}~=~{1 \over 2}\{[2b^{\prime\prime}(\overline{3})_{3}+c^{\prime\prime}(\overline{3})_{3}+2a^{\prime\prime}(6)_{3}+b^{\prime\prime}(6)_{1}-d^{\prime\prime}(6)_{2}  \nonumber\\
&&\qquad\qquad
+6a^{\prime\prime}(\overline{15})_{3}+3b^{\prime\prime}(\overline{15})_{1}+c^{\prime\prime}(\overline{15})_{3}-c^{\prime\prime}(\overline{15})_{2}+3d^{\prime\prime}(\overline{15})_{2}
]\nonumber\\
&&\qquad\qquad+[b^{\prime\prime}(\overline{3})_{1}+b^{\prime\prime}(\overline{3})_{2}+c^{\prime\prime}(\overline{3})_{1}+a^{\prime\prime}(6)_{1}+a^{\prime\prime}(6)_{2}+b^{\prime\prime}(6)_{3}-c^{\prime\prime}(6)_{1}-c^{\prime\prime}(6)_{3}
\nonumber\\
&&\qquad\qquad
-d^{\prime\prime}(6)_{3}+3a^{\prime\prime}(\overline{15})_{1}+3a^{\prime\prime}(\overline{15})_{2}+3b^{\prime\prime}(\overline{15})_{3}+3c^{\prime\prime}(\overline{15})_{1}-c^{\prime\prime}(\overline{15})_{2}+3d^{\prime\prime}(\overline{15})_{3}
] \}\nonumber\\
\end{eqnarray}

\end{document}